\documentclass[manuscript]{aastex}

\shortauthors{Kafka et al.}

\begin{document}

\title{QU Carinae: Supernova Ia in the making?}


\author{S. Kafka\altaffilmark{1}}
\affil{Department of Terrestrial Magnetism, Carnegie Institution of Washington, 5241 Broad Branch Road NW, Washington, DC 20015, USA }

\author{R. K. Honeycutt\altaffilmark{2}}
\affil{Astronomy Department, Indiana University, Swain Hall West, Bloomington, IN 47405, USA}

\author{R. Williams\altaffilmark{3}}
\affil{Space Telescope Science Institute, 3700 San Martin Drive Baltimore, MD 21218, USA}




\begin{abstract}

Variable NaI absorption lines have been reported in a number of type Ia supernovae (SNeIa). The presence of this circumstellar material suggests that  cataclysmic variables (CVs) with a giant donor star may be the progenitors of these SNeIa (Patat et al. 2007). We present echelle spectra of the CV QU Carinae which strengthen the connection between CVs of the V Sge class, the Accretion Wind Evolution scenario, variable wind features, variable NaI absorption, and SNIa. This thread not only provides insight into the spectral peculiarities of QU Car, but also links SNeIa as a class with their parent systems.

\end{abstract}


\section{Introduction}

Although the progenitors of Supernovae Ia (SNeIa) are still controversial, it is accepted that, at least some of the events, originate from detached or semi-detached binary star systems in which at least one of the two components is a massive C-O white dwarf (WD). In semi-detached systems, either via Roche lobe overflow of the companion or via a wind, the WD accumulates H or He-rich material which is then burned to C and O. Under the right conditions (which are primarily controlled by the stability of the mass-accretion rate onto the WD) the net WD mass reaches the Chandrasekhar mass limit initiating a series of thermonuclear reactions eventually leading to a SNeIa. Although outlining this scenario seems to be rather straightforward, its specifics are far from being understood or defined, especially the nature of the mass-losing star.

Among the most promising SNeIa progenitors are the V Sagittae (V Sge)-type cataclysmic variables (CVs) \citep{1998PASP..110..276S}. These are semi-detached binaries consisting of a WD and a giant or main sequence donor star, with high mass transfer rates ($\sim$10$^{-7}$-10$^{-5}$M$_{\sun}$/yr). This rate allows for stable nuclear burning on the surface of the WD, which can take place when a relatively massive (0.7-1.2 M$_\sun$; van den Heuvel et al. 1992) WD accretes near Eddington rates ($\sim$10$^{-6} M_{\sun}$/ yr). 

The process of a WD accumulating mass in a V~Sge-type system, gradually reaching the critical Chadrasekhar mass, is described in detail by the accretion wind evolution scenario (AWE), coined by Hachisu \& Kato (2003a and references therein). AWE successfully reproduces the long-term light curve of the prototype of the category, V~Sge and its Large Magellanic Cloud twin RX~J0513.9-6951 (Hachisu \& Kato 2003b) accounting for the bright and faint states of the systems and the transitions between them. When the accumulated envelope on the WD reaches a critical mass, the WD atmosphere expands, generating a massive wind with $\dot{M}_{wind}\sim$10$^{-7}$M$_{\sun}$yr$^{-1}$.  This wind drives a Kelvin-Helmholtz instability at its interface with the disk, peeling off the surface layer of the disk, which completely obscures soft x-rays from escaping from the binary. With time (and accumulated material) $\dot{M}_{wind}$ increases reaching $\sim$10$^{-5}$M$_{\sun}$yr$^{-1}$, in which case the mass of the WD envelope is further reduced by wind mass losses. The chromosphere of the companion is then eroded by the wind to the point where the donor star shrinks inside its Roche lobe and accretion (through L1) temporarily stops. The wind mass loss is then decreased, eventually reaching zero, and the system transitions to a faint state. During this faint state, soft x-rays emerge. The system gradually recovers to the bright state when the donor star regains contact with its Roche lobe and accretion commences; the cycle starts again.

Other than the prototype, there are 4 suspected members in this category \citep{1998PASP..110..276S}, but other than V Sge itself, their properties are not well-known. Recently, the Galactic CV QU Carinae (hereafter QU Car) was suggested to also be a V Sge star (Kafka et al. 2008--hereafter Paper~I). The system displays strong and pronounced outflows in the UV and in optical Carbon lines, reaching velocities of at least 5700km s$^{-1}$ (Drew et al. 2003 and Paper~I). At the same time, strong [OIII]~5007 emisson confirms the presence of a circumbinary nebula (Paper~I). QU Car, being the brightest member in the V~Sge category, may be a galactic twin of supersoft X-ray sources. Here, we present new high resolution spectra of QU Car revealing the highly variable nature of the system on timescales of years, and we discuss its SNeIa progenitor status.

\section{Observations}

Spectra were obtained with the Echelle Spectrograph on the du Pont 2.5-m telescope of the Las Campanas Observatory\footnote{http://www.lco.cl/} during 6 nights of observations. The Echelle Spectrograph provides wavelength coverage from 3700-9000$\AA$ at a typical resolution of $\sim$30,000 (using the 1'' slit). Spectra of a ThAr lamp were obtained for wavelength calibration at the position of the telescope, just before and just after each stellar exposure. A short log of observations is presented in table~1. We use the designations E1-E6 for each of the six observing epochs and will refer to them as such for the remainder of this work. With this setup, we obtained a total of 14 spectra of QU~Car. For data processing and reductions we used IRAF's\footnote{IRAF is distributed by the National  Optical Astronomy Observatories, which are operated by the Association of  Universities for Research in Astronomy, Inc., under cooperative agreement with the National Science Foundation.} echelle package. Furthermore, radial velocity standard stars were observed each night, and were used to confirm that our velocities for spectral features in QU Car had no systematic errors. No spectrophotometric or telluric standards were obtained during our observing runs, therefore telluric features are present in the reduced spectra. All velocities discussed in this paper are heliocentric.

The weather was clear for most of our run. Passing clouds compromised the S/N of the E4 spectra despite our increased exposure time; therefore we will not use the E4 spectra in our analysis. 

\section{Discussion}

\begin{figure}
\includegraphics[width=10cm, angle=0.0]{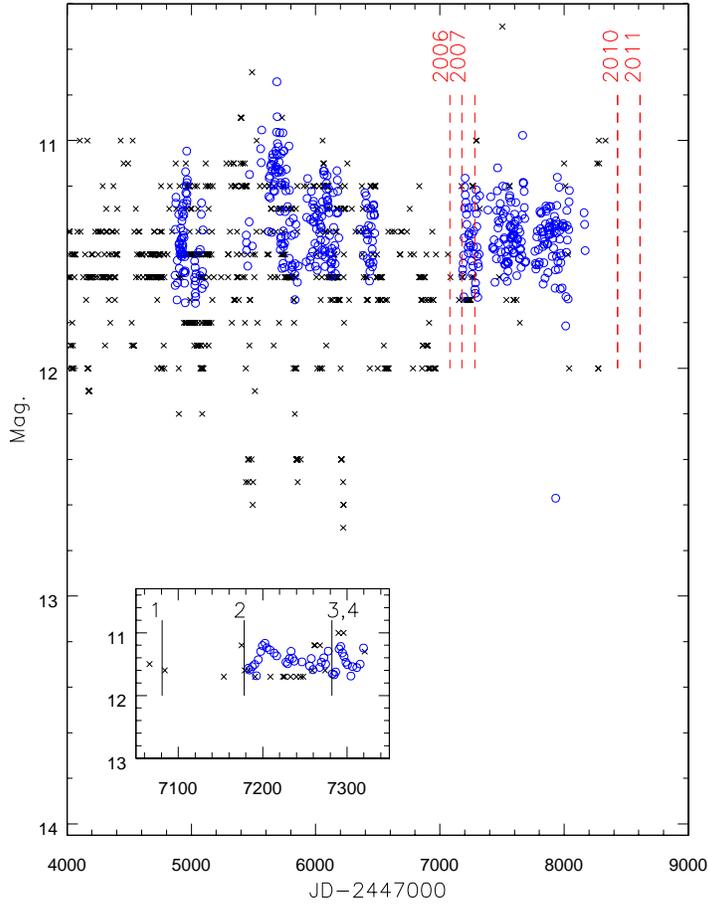}
\caption{Visual AAVSO and V Mag. ASAS light curves of QU Car. This
  plot demonstrates that bright and faint states are present in the
  long-term optical behavior of QU Car, although the transitions are
  not well defined due to lack of continuous monitoring. The four epochs of the 2006-2007 low-resolution spectroscopic observations of paper~1  are highlighted in the inset, to demonstrate that QU Car was at its  ``bright'' magnitude at the time of the relevant
  observations. Photometric data were not available during our 2010-2011 observations.
\label{lc}}
\end{figure}

In V Sge, the optical photometric bright state lasts for about 170 days and the faint state for about 130 days (Robertson et al. 1997; \citep{1999A&AS..139...75S}. Figure~1 presents part of the long-term light curve of QU Car: visual data are from the database of the American Association of Variable Star Observers (AAVSO)\footnote{American Association of Variable Star Observers (AAVSO), Henden, A.A., 2009, Observations from the AAVSO International Database, private communication. http://www.aavso.org} and the V magnitude data are from the All Sky Automated Survey (ASAS)\footnote{All Sky Automated Survey (ASAS)  http://www.astrouw.edu.pl/asas, Pojmanski (1997).}. The full (more than 20 years) AAVSO visual light curve is also presented and discussed in Paper~I. The ASAS V-mag light curve has a detection limit of V$\sim$14 mag. Sporadic coverage does not allow for a detailed study of the short-term behavior of the system, however the light curves do indicate the occasional presence of ``faint'' states, where the brightness drops below visual 12 mag (also see Paper~I). Albeit short, those faint states can last for $\sim$100~days before the system recovers to its usual V ``bright'' magnitude of $\sim$11.5. The inset in figure~\ref{lc} is a closeup of the light curves at the times of our 2006/2007 observations, also presented in Paper~I, where the system was in its bright V$\sim$11.5 magnitude, exhibiting low-amplitude erratic variations which are normal for nova-like CVs. No light curves exist for the times of the echelle data of 2010/2011. Therefore, we will use the spectral appearance of QU Car from (Paper~I) as a guide on the behavior of the system in its bright state and as a comparison for the new spectroscopic data. Overall, the characteristics of the bright states of QU Car have been explored spectroscopically in \citep{1982ApJ...261..617G}, in  \citep{2003MNRAS.338..401D}, and in Paper~I.

As discussed in Paper~I, in the bright state the system exhibits variable and irregular Balmer emission lines, HeI and HeII emission and one of the strongest Bowen blend (CIII/NIII/OII at 4640-4650$\AA$) among all CVs. The broad CIV 5807 $\AA$ emission is accompanied by pronounced P-Cygni profiles reaching velocities of $\sim$5700 km/sec. Excess of CNO-processed material in the optical spectra likely originates from a wind expelled from the binary's C-O WD, in agreement with the AWE scenario and with relevant UV observations \citep{2003MNRAS.338..401D}. 
 It is critical to stress that such a wind, carrying away the outer layers of the disk and of the donor star, is necessary in the AWE scenario in order for the WD to eventually reach the Chandrasekhar mass; otherwise the WD envelope expands and forms a circumbinary envelope, eventually leading to merging of the two stellar components instead of a SNeIa \citep{2008ApJ...679.1390H}.
The presence of [OIII] 5007$\AA$ and [NII] 6584$\AA$ at all four epochs of the bright state in Paper~I, is the imprint of a nebula in the vicinity of the binary and a relic of material expelled from it.

In following sections we will discuss the time variability of the QU Car
spectrum in our high resolution data.

 Gilliland \& Phillips (1982) provided an orbital period of 10.9 hr for QU Car, however this period was not confirmed in Paper I and there is not yet a reliable
 ephemeris for the binary. The spectral variability discussed in this paper may be due to the varying viewing angle of the different line-forming
 regions as the system revolves.  On the other hand the changes may be due to variations in the strength and motion of the wind. We argue for the latter interpretation, but we admit that the lack of an ephemeris precludes our testing for orbital dependences.

Since we don't have flux-calibrated echelle spectra in hand, we use IRAF's {\it continuum} function to normalize the spectra in each echelle order. This allows comparison of the spectra at different epochs, however it smeared out features that can be as wide as the echelle orders themselves. In following subsections we will discuss various emission and absorption features seperately.

Finally, to address the origin of various components, we need to assess the accuracy with which we can measure the radial velocities of the lines and the possible systematic errors affecting their respective values. It is always possible that instrumental effects (echelle spectrograph flexure) introduce pixel shifts and pseudo-variations in the radial velocities of the lines. Taking comparison lamp spectra before and after each object exposure is a common procedure to minimize this effect; nevertheless, an independent confirmation is necessary. In the spectrum of QU~Car, there are two diffuse interstellar bands (DIBs) at 5870$\AA$ and 5797$\AA$  at adjacent blue orders, the oxygen A and B bands and the telluric bands at red orders. We have used the narrow DIB features to ascertain the accuracy of our wavelength calibrations inasmuch as the radial velocity shifts in the Na I D absorption occurs at different temporal epochs and could conceivably be due to slight changes in the wavelength calibration due to the instrument configuration.  Measurement of the DIBs radial velocities confirms the accuracy of our radial velocity determinations within an RMS error of $\sim$$\pm$0.10 km/sec.

\subsection{Emission lines}

Table 2 lists the equivalent widths (EWs) of the most prominent features
in our echelle spectra. By comparing to the emission line strengths in 2006/2007 (Table 3 of Paper~I) we see that the new EWs are in general consistent with those measured in 2006/2007 using
 lower resolution data.  Both the 2006/2007 EWs and the new data are characterized
 by occasional changes of over 2$\times$ between nights, and sometimes
 within
 a night, mixed with intervals of relative stability.  Note in particular
 that
 in 2006/2007 the Bowen blend was nearly constant over 6 months at
 EW$\sim$3.2\AA,
 which is a little weaker than in 1979/1980 (Gilliland \& Phillips 1982). 
 In our
 2010 spectra this feature is much weaker, and has practically disappeared
 by
 E3.  The Balmer lines are also weak at this time, but HeII is stronger
 than the 2006-2011 average.  Apparently these various emission lines originate in quite
 different
 regions of the system, which can vary independently.

\begin{figure}
\includegraphics[width=12cm, angle=0.0]{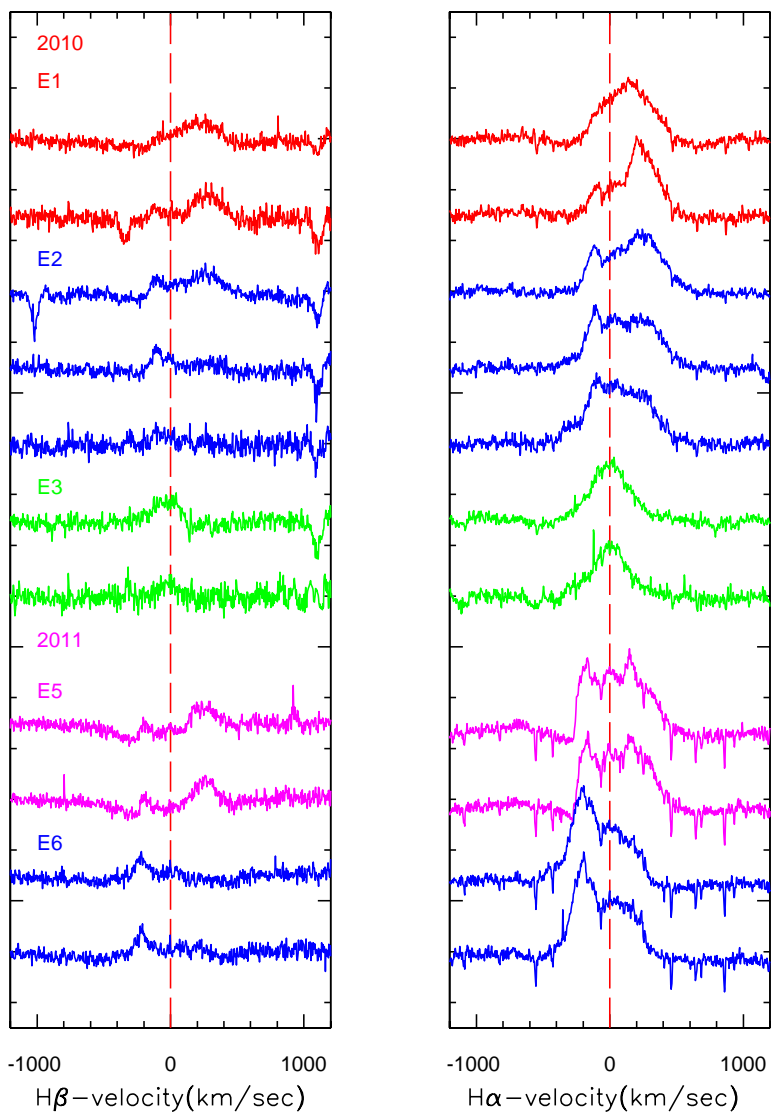}
\caption{Balmer lines for epochs E1-E6 (E4 is missing, since the
  relevant data are not used in this work). The Balmer line profiles
  seem to be smoother in 2010, with multiple absorption components in
  2011. A stationary redshifted emission component at $\sim$-135km/sec
  is representing stationary circumbinary gas. Finally, the
  blueshifted and redshifted absorption troughs in the 2010 H$\beta$
  line likely represents material expelled from the binary. \label{balmer}}
\end{figure}

\begin{figure}
\includegraphics[width=12cm, angle=0.0]{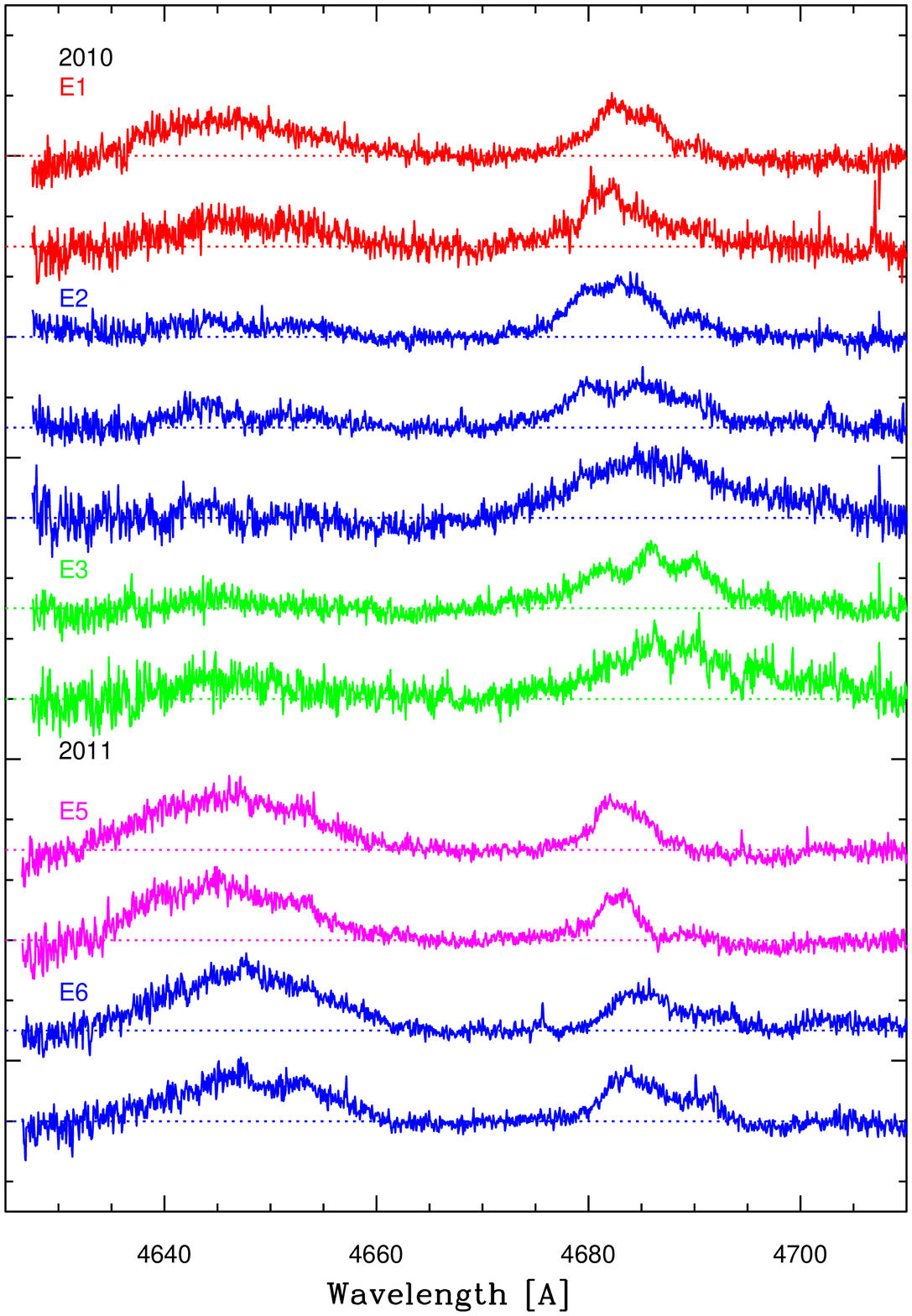}
\caption{Same as in figure~\ref{balmer}, but for the Bowen blend and
  HeII4686 emission. The dashed horizontal lines guide
  the eye to the normalized continuum. Notice the attenuated Bowen
  blend lines in 2010 with respect to their appearance in 2011.\label{bowen}}
\end{figure}

Figures 2 and 3 show the line profiles of some of the emission lines,
 where the higher spectral resolution of the echelle data reveal considerable details compared to earlier studies.  At times the H$\alpha$ profile resembles the line profiles in V Sge (Gies, Shafter \& Wiggs 1998; Robertson, Honeycutt \& Pier 1997), with
 multiple components whose relative strengths vary.  Note that the H$\alpha$
 profile
 sometimes shows weak blue-shifted absorption at a velocity of $\sim$-300
 km s$^{-1}$
 This wind-induced P-Cygni feature was also seen in the H$\alpha$ profiles of
 Paper I.
 
 For all four of the 2011 spectra (E5 and E6) the H$\alpha$ emission line is
 accompanied by 7 narrow weak absorption features.  These features are
 identical
 in these 4 spectra (separated by 3 nights), and do not appear in any of
 the
 2010 spectra (E1-E3).  They do not agree with the wavelengths of telluric
 water features (e.g. Vince \& Vince 2010) nor do they appear in our standard stars (or any of the stars observed the same night in neighboring parts of the sky, bracketing the QU Car observations). In the absence of an accurate ephemeris, we have few clues to the physical locations of those lines.

The fact that the EWs of HeII and HeI do not follow changes in the Balmer 
and Bowen blend suggests that those lines have different origins in the system (as expected from their different excitation potentials). Using radial velocity arguments Gilliland and Phillips (1982) argue that CIII~4650$\AA$ should be the dominant species in the Bowen blend. Furthermore, UV spectra \citep{2003MNRAS.338..401D} indicate that carbon is overabundant in QU Car. A Carbon overabundance in QU Car is also supported by the presence of the strong CIV~5801,5812$\AA$ emission, in the Gilliland \& Phillips (1982), Drew et al. (2003) and Paper~I optical spectra.  Drew et al. (2003) argue that this carbon enhancement should originate from the envelope of the donor star, suggesting that the star is an early-R type. However, the observed variations in the Bowen blend and in CIV in our spectra argue against such an interpretation. Emission from the secondary star could not have been diminished in one epoch of our observations but still be present at a different one. In 2010 both the Bowen blend and CIV are significantly reduced in strength; this is an indication that the lines are variable in nature. A plaussible interpretation is that carbon originates from the atmosphere of the the white dwarf of the binary, which is attenuated in the faint state. In 2011, when the CIII/NIII/OII line strength of the binary returned to its ``bright'' state value, weak CIV~5801,5812$\AA$ emission is also present.

In our 2006/2007 spectra, we also detected variable [OIII]~5007$\AA$ and [NII]~6584$\AA$ emission, indicative of the presence of a nebula around the system. The [OIII]~5007$\AA$ lines appeared to have two components, ranging in velocities from -500km/sec to 370km/sec - probably representing the front and back of an expanding shell or an outflow. In our new echelle data, only one (weak) component of the [OIII]~5007$\AA$ line is present; its strength is reduced with time (perhaps with orbital phase) between E1 and E2. In 2011, the [OIII]~5007$\AA$ emission is present only during E5, with no accompanying [NII]~6584$\AA$ emission. This erratic variation in the oxygen emission line strength does not seem to be associated with variations in the strength of the Bowen blend. However, we are in need of a better ephemeris for the system in order to correlate the observed oxygen emission to a specific location and process on the binary.

\subsection{H$\beta$ components: evidence for circumstellar material?}

We now turn our attention to the mysterious absorption components bracketing H$\beta$ in 2010. Having excluded contamination in our spectra (flat fielding artifacts, data reduction faux pas or background reduction residuals), the features are convincingly real. Since there are no strong telluric features at this part of the spectrum (nor is there any chance for this absorption to come from the interstellar medium, especially considering its variable nature and appearance), the features should originate from the binary. The red component is stationary at $\sim$1110km/sec. The blue component is present only during two subsequent nights of our observations (second spectrum in E1 and first spectrum in E2, at -370km/sec and -1050km/sec respectivelly). Considering that our exposure times are 2000 sec, we are surprised to not detect this red component in all spectra in E1 and in E2. 

Among CVs, those features are unique to QU Car.  However, similar absorption lines sometimes appear in other accreting sources: complex blueshifted and redshifted absorption components emerge in the Balmer lines of Herbig Ae/Be stars (e.g. Guimaraes et al. 2006). In this case, the redshifted absorption is usually interpreted as being due to material accreted onto the star from its inner disk (sometimes via magnetic fields) and the blueshifted one as being due to mass loss (outflows). Both processes can happen simultaneously (e.g. Natta, et al. 2000). Taking the Ae/Be work as a starting point, we could attribute the blue components to ejecta from QU Car, while the redshifted components originate on an inflow from the cirumstellar medium. It is difficult to imagine a mechanism that allows for both processes simultaneously, unless the flows are magnetically controlled (as is the case in many T Tauri stars).  However, there is no evidence that QU Car contains a highly magnetic white dwarf  (i.e., no Zeeman or cyclotron features). 
Furthermore, it seems unlikely that this material represents a jet. Jet-like features in CVs sometimes appear in dwarf novae in outburst or in some novae, both times representing material ejected from the binary during explosions (e.g. Cowley et al. 1998). In semi-detached binaries, optical jets also appear in symbiotic stars and in low mass x-ray binaries as {\it emission} components moving away from Balmer lines with velocities reaching 4000km/sec ( e.g.  Cowley et al. 1998). 
Perhaps the gas motions in QU Car are so turbulent and complex that portions of the flow can be both red and blue shifted in front of the same continuum source. It is unfortunate that we lack optical photometry (or any light curves) at the time of our spectroscopic observations (stressing again the importance of having simultaneous multiwavelength observations for a comprehensive understanding of QU Car).

\subsection{ NaI D absorption: connection with the supernovae Ia?}

The unusual appearance of the NaID lines is revealed in figure~4: in 2010, each component consists of weak blueshifted emission accompanied by redshifted absorption! Those lines (and the adjacent HeI 5876 emission) are at the center of the order during all our observing runs, ruling out radial velocity variations introduced by pixel shifts due to poor echelle order distortion correction at the edges of each order. The EW and RV values are given in Table~3, where missing entries indicate 
that the line was not detected. The emission components are very narrow and appear slightly blueshifted with respect to the line's rest wavelength. Furthermore, this emission is quite variable and present only during 2010. No correlation of the line strength and radial velocity was found. In CVs, the NaID lines are usually in absorption, arising from the photosphere of the donor star. Sodium is usually ionized in accretion disks; its presence in emission in QU Car is indicative of lower than usual temperatures. This suggests a different optical state for the system, perhaps similar to V Sge's "faint" state, where the mass transfer rate is significant lower, in agreement with the reduced strength of the Balmer and the Bowen blend line profiles. We are intrigued by the "stationary" (constant velocity) nature of this NaID emission component over the four nights of our observations in 2010. {\em If} this line has an origin on the accretion disk or on the hotspot, it may indicate that the QU Car's disk outer edge reaches the center of mass of the binary, in agreement with QU Car being an unusually bright object.

The RVs of both NaD absorption components are constant at ~$\sim$-7 km s$^{-1}$ during E1-E3.  Eight months
later these RVs are again constant but at ~$\sim$-13 km s$^{-1}$ during E5-E6. Both components of the NaID are well displaced from their rest wavelength as they accelerate between the 8 months of observations. 
In the bright state, variable NaID is also present (e.g. Paper~I), however, with the lower resolution of Paper~I it was not possible to resolve individual absorption components. Furthermore, in our echelle spectra, we cannot kinematically associate any other emission line components with the absorption features. 

Since neither component of the NaID doublet is at the local standard of rest velocity, we explore the possibility that they are interstellar in origin, following the Galactic rotation at the location of QU Car.  \citep{2003MNRAS.338..401D} estimate the galactic rotation velocity at $\sim$2kpc along the line-of-sight to QU Car to be -18.3km/sec and attribute FeII~1608$\AA$ and SII~1250$\AA$ bluehifted line components to interstellar absorption. As a side-note, the minimum distance to QU Car is suggested to be 500pc (Linnell et al. 2008 and references therein), which would imply less interstellar absorption. However, for our purposes, the distance to QU Car and amount of interstellar abosprtion is irrelevant since we are discussing radial velocity {\it variations} in the absorption lines, not their absolute values.\footnote{ It is also possible that an interstellar component is present and unresolved in the observed lines, however it will simply introduce an error in the estimate of the line EW.} In this case, the variable velocities of NaID components between 2010 and 2011 argue against an interstellar origin. Moreover, NaID absorption could originate on the donor star of the binary. Because of the absence of other photospheric identification features from this star (e.g. FeI, CaI, MgII, Ti or even Balmer absorption) we rule out this possibility. In CVs NaI absorption lines can be generated in nova explosions (e.g. Shore et al. 2011); however QU Car has no indications of recent erruptions.

A possible scenario comes from the supernova community: \citep{2007Sci...317..924P} describes variable NaI D absorption in the vicinity of SN 2006X, originating from circumbinary clouds, heated by the SN and interacting with the explosion ejecta. Patat et al. (2007) argue that the NaI was present in the SNeIa progenitor as circumbinary material. UV radiation from the SNeIa explosion ionized this material, which slowly recombined post-SNeIa to produce the observed variable NaID absorption. The absence of CaII H\&K absorption lines was attributed to the lack a radiation field hot enought to significantly ionize Ca.

Variable NaID absorption has now been detected in a group of SNeIa (Sternberg et al. 2011; Simon et al. 2009; Simon et al.2007), favoring the single degenerate scenario (C-O WD + giant donor star) for a SNeIa progenitor. Circumstellar clouds continuously replenished by a wind from the giant donor star provide a plausible mechanism for the formation of NaID lines in QU Car. Therefore, our detection of variable NaID lines around QU~Car provides, for the first time, a bridge between this class of SNe and their elusive progenitors. 

\begin{figure}
\includegraphics[width=10cm, angle=0.0]{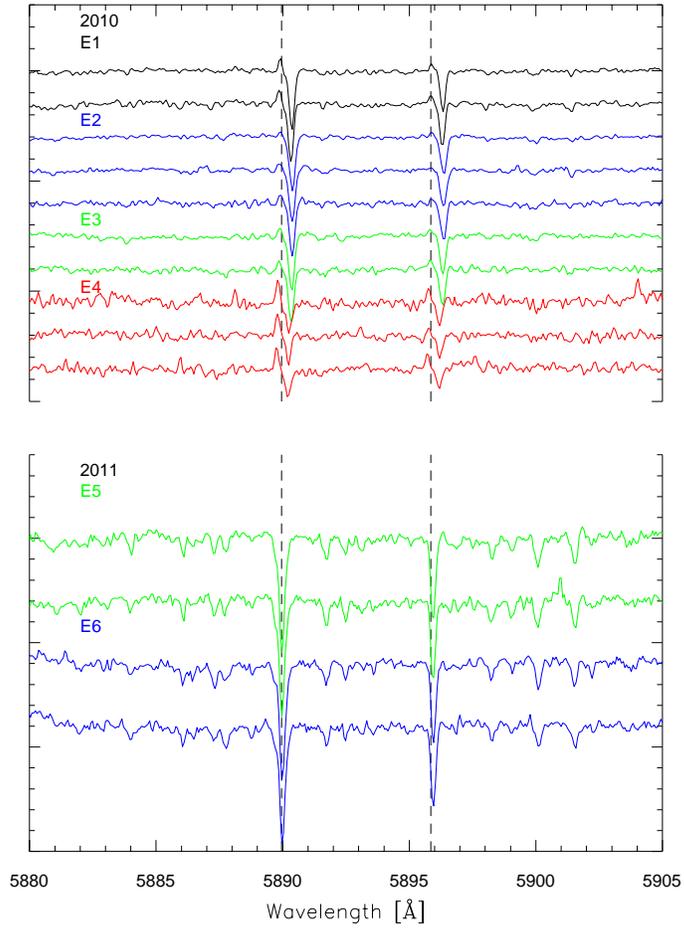}
\caption{Same as figure~3, for the NaD line region.}
\end{figure}

\section{Final remarks}

Williams et al. (2008) has reported variable Transient Heavy Element Absorption (THEA) lines after a nova explosion, likely originating from interactions between circumbinary material and the nova ejecta. Density enhancements of circumbinary material have also been detected around the recurrent nova RS Oph (Iijima~2008; Patat et  al.~2011) likely originating from expanding material from the donor star, urging observers to conduct similar studies. QU Car is the first system in which such material is discovered and characterized, providing evidence that high $\dot{M}$ CVs are the likely progenitors of SNeIa. We have presented two epochs of high-resolution observations of this system, attributing emission from the carbon-dominated Bowen blend to the white dwarf of the binary. Strong and variable absorption components to the H$\beta$ emission lines attest to the presence of circumbinary material that is receding from the binary. Furthermore, the presence of blueshifted NaID absorption components are interpreted as being due to circumbinary material likely removed from the donor star by a strong wind. Such material has also been detected around SNeIa, and is now a signature of the donor star of some SNeIa. Most importantly, this material corroborates the validity of the accretion wind evolution scenario for mass increase onto the white dwarf, eventually leading to a SNeIa explosion.

The variability of the spectrum of
 QU Car presents many mysteries and challenges, ranging from the
 unidentified absorption
 features near the Balmer lines, to the multiple components of the strong
 emission
 lines, to the redshifted NaI~D absorption line profiles with blue emission
 components.  To add to the confusion, most of these effects are highly
 variable.
 It is frustrating that our attempts to understand these phenomena have not
 been
 broadly successful, but we do wish to draw attention to these features in
 order to
 encourage further work.   Also, we wish to emphasize that the lack of a
 reliable orbital ephemeris greatly complicates the interpretation of
 widely-spaced spectral features such as the ones presented here.  
Our present study documents for the first time, a phase where accretion is diminished in this system in 2010. More importantly, this new study ties together the chasm between observations of SNeIa and their likely progenitors, providing a viable candidate of a SNe in the making. 
QU Car is indeed a unique system, even among famously divergent CVs. 

\section*{Acknowledgments}
We would like to thank our referee, Dr Steiner, for the careful review of our manuscript. We acknowledge with thanks the variable star observations from the AAVSO International Database contributed by observers worldwide and used in this research. SK also acknowledges support by the NASA Astrobiology Institute.


\clearpage


\begin{table*}
 \centering
 \begin{minipage}{140mm}
  \caption{Log of observations}
  \begin{tabular}{cccc}
\hline
Date of obs. & No observation & exp. time (s) & Notes \\
\hline
E1: 2010-08-19  &  2 & 1800 & Clear\\
E2: 2010-08-20  &  3 & 1800 & Clear\\
E3: 2010-08-21  &  2 & 2000 & Clear\\
E4: 2010-08-24  &  3 & 2000/2400 &  Partly cloudy\\
E5: 2011-02-16  &  2 & 1500 & Clear\\
E6: 2011-02-19  &  2 & 1500 & Clear\\
\hline
\end{tabular}
\end{minipage}
\end{table*}


\begin{table*}
 \centering
 \begin{minipage}{140mm}
  \caption{EWs (in \AA) of the emission features of QU Car. The EW measured error is 0.02\AA}
  \begin{tabular}{ccrccccccc}
  \hline
Date obs. & Epoch & HJD-2455428 & Bowen & HeII & H$\beta$ & HeI & H$\alpha$ & HeI & HeI \\
&  &  & CIII/NIII/OII & 4686 $\AA$ &  & 5876 $\AA$ & & 6678 $\AA$ & 7065 $\AA$\\
\hline
2010 &E1  &  0.501  &  -2.23  &  -1.47  &  -0.77  &  -0.18  &  -2.72  &  -1.12  &  -0.38  \\
         &      &  0.556  &  -1.78  &  -1.98  &  -1.25  &  -0.85  &  -3.11  &  -1.38  &  -0.36  \\
         &E2  &  1.399  &  -0.70  &  -1.89  &  -0.79  &  -0.10  &  -3.23  &  -0.73  &  -1.13  \\
         &      &  1.546  &  -1.00  &  -2.13  &  -0.61  &  -0.39  &  -3.78  &  -0.12  &  -1.20  \\
         &      &  1.526  &  -1.03  &  -5.14  &  ...  &  -0.39  &  -2.98  &  -0.58  &  -1.62  \\
         &E3  &  2.479  &  -0.06  &  -2.75  &  -0.63  &  -0.99  &  -2.26  &  ...  &  ...  \\
         &      &  2.503  &  -1.00  &  -3.05  &  -0.28  &  ...  &  -2.28  &  ...  &  ...  \\
 2011 & E5 &  180.757  &  -3.39  &  -1.03  &  -0.57  &  -0.44  &  -4.07  &  -0.66  &  ...  \\
         &      &  180.775  &  -3.31  &  -0.74  &  -0.69  &  -0.24  &  -3.01  &  -0.80  &  ...  \\
         &E6  &  183.831  &  -3.42  &  -1.08  &  -1.43  &  -0.26  &  -3.75  &  -1.42  &  -0.98  \\
         &      &  183.849  &  -2.34  &  -1.35  &  -1.01  &  -0.38  &  -4.47  &  -1.18  &  -1.16  \\
\hline
\end{tabular}
\end{minipage}
\end{table*}


\begin{table*}
 \centering
 \begin{minipage}{140mm}
  \caption{RV (in km/sec (RV+/-0.10km/sec) and EW (in \AA+/-0.02 \AA) measurements of the NaD lines}
  \begin{tabular}{crcccccc}
  \hline
\hline
 & HJD-2455428   & & RV (km/sec) & EW $\AA$  & & RV (km/sec) & EW $\AA$ \\ 
\hline
      &          & &NaD1 emission & NaD1 emission && NaD1 absorption & NaD1 absorption \\
\hline
E1    &  0.501   &  &  16.66    &  -0.02  &&  -6.98   &  0.14   \\
      &  0.556   &  &  18.73    &  -0.03  &&  -5.01   &  0.15   \\
E2    &  1.400   &  &  16.26    &  -0.01  &&  -7.93   &  0.14   \\
      &  1.546   &  &  ...      &...      &&  -7.52   &  0.13   \\
      &  1.526   &  &  ...      & ...     &&  -7.62   &  0.11   \\
E3    &  2.479   &  &  17.42    &  -0.01  &&  -6.72   &  0.14   \\
      &  2.503   &  &  18.90    &  -0.02  &&  -5.80   &  0.12   \\
E5    &  180.757 &  &  ...      &  ...    &&  -13.37  &  0.15   \\
      &  180.775 &  &  ...      &  ...    &&  -13.15  &  0.16   \\
E6    &  183.831 &  &  ...      &  ...    && -13.71   &  0.16   \\
      &  183.849 &  &  ...      &  ...    && -13.76   &  0.16   \\
\hline
      &          & &NaD2 emission & NaD2 emission && NaD2 absorption & NaD2 absorption \\
\hline
E1    &  0.501    &  &  16.30   &  -0.02  && -6.75     &  0.09  \\
      &  0.556    &  & 17.75    &  -0.02  &&  -5.50    &  0.09  \\
E2    &  1.400    &  &  15.80   &  -0.01  &&  -8.52    &  0.10  \\
      &  1.546    &  & ...      &  ...    &&  -7.96    &  0.10  \\
      &  1.526    &  & ...      &  ...    &&  -8.05    &  0.09  \\
E3    &  2.479    &  & 18.79    &  -0.02  &&  -6.75    &  0.09  \\
      &  2.503    &  &  18.84   &  -0.03  &&  -6.04    &  0.09  \\
E5    &  180.757  &  &  ...     &  ...    &&  -13.22   &  0.09  \\
      &  180.775  &  &  ...     &  ...    &&  -13.25   &  0.08  \\
E6    &  183.831  &  &  ...     &  ...    &&  -13.77   &  0.08  \\
      &  183.849  &  &  ...     &  ...    &&  -13.40   &  0.09  \\
\hline
\end{tabular}
\end{minipage}
\end{table*}

\end{document}